\documentclass[prb,nofootinbib,notitlepage,longbibliography,superscriptaddress,twocolumn]{revtex4-2}
\usepackage[pdftex]{graphicx}
\usepackage[greek,english]{babel}
\usepackage{lmodern}
\usepackage{bbold}
\usepackage{enumerate}
\usepackage{colortbl}
\usepackage{dcolumn}
\usepackage{makeidx}
\usepackage{hyperref}
\usepackage{float}
\usepackage{braket}
\usepackage{physics}
\usepackage{amsmath,amsfonts}
\usepackage{mathtools}
\usepackage{ytableau}
\usepackage{soul}
\usepackage{comment}
\usepackage[compat=1.1.0]{tikz-feynman}

\newcommand{\MM}{\overline{\mathbf{M}}}
\newcommand{\bl}{{\boldsymbol{\lambda}}}
\newcommand{\br}{{\boldsymbol{\rho}}}
\newcommand{\bg}{{\boldsymbol{\gamma}}}
\newcommand{\bog}{{\overline{\boldsymbol{\gamma}}}}

\makeatletter
\DeclareFontFamily{OMX}{MnSymbolE}{}
\DeclareSymbolFont{MnLargeSymbols}{OMX}{MnSymbolE}{m}{n}
\SetSymbolFont{MnLargeSymbols}{bold}{OMX}{MnSymbolE}{b}{n}
\DeclareFontShape{OMX}{MnSymbolE}{m}{n}{
    <-6>  MnSymbolE5
   <6-7>  MnSymbolE6
   <7-8>  MnSymbolE7
   <8-9>  MnSymbolE8
   <9-10> MnSymbolE9
  <10-12> MnSymbolE10
  <12->   MnSymbolE12
}{}
\DeclareFontShape{OMX}{MnSymbolE}{b}{n}{
    <-6>  MnSymbolE-Bold5
   <6-7>  MnSymbolE-Bold6
   <7-8>  MnSymbolE-Bold7
   <8-9>  MnSymbolE-Bold8
   <9-10> MnSymbolE-Bold9
  <10-12> MnSymbolE-Bold10
  <12->   MnSymbolE-Bold12
}{}

\let\llangle\@undefined
\let\rrangle\@undefined
\DeclareMathDelimiter{\llangle}{\mathopen}%
                     {MnLargeSymbols}{'164}{MnLargeSymbols}{'164}
\DeclareMathDelimiter{\rrangle}{\mathclose}%
                     {MnLargeSymbols}{'171}{MnLargeSymbols}{'171}
\makeatother

\begin{document}
\title{Bipartite Sachdev-Ye Models with Read-Saleur Symmetries}
\author{J. Classen-Howes}
\affiliation{Rudolf Peierls  Centre  for  Theoretical  Physics, Parks Road, Oxford  OX1  3PU,  United  Kingdom}
\author{P. Fendley}
\affiliation{All Souls College, University of Oxford, OX1 4AL}
\affiliation{Rudolf Peierls  Centre  for  Theoretical  Physics, Parks Road, Oxford  OX1  3PU,  United  Kingdom}
\author{A. Pandey}
\affiliation{Department of Physics, Stanford University, Stanford, CA 93405, USA}
\author{S. A. Parameswaran}
\affiliation{Rudolf Peierls  Centre  for  Theoretical  Physics, Parks Road, Oxford  OX1  3PU,  United  Kingdom}
\begin{abstract}
We introduce an SU($M$)-symmetric disordered bipartite 
spin model with unusual characteristics. Although superficially similar to the Sachdev-Ye model, it has several markedly different properties for $M\geq3$. In particular, it has a large non-trivial nullspace whose dimension grows exponentially with system size. The states in this nullspace are frustration-free, and are ground states when the interactions are ferromagnetic. The exponential growth of the nullspace leads to Hilbert-space fragmentation and a violation of the eigenstate thermalization hypothesis. We demonstrate that the commutant algebra responsible for this fragmentation is a non-trivial subalgebra of the Read-Saleur commutant algebra of certain nearest-neighbour models such as the spin-1 biquadratic spin chain. 
We also discuss the low-energy behaviour of correlations for the disordered version of this model in the limit of a large number of spins and large $M$, using techniques similar to those applied to the SY model.
We conclude by generalizing the Shiraishi-Mori embedding formalism to non-local models, and apply it to turn some of our nullspace states into quantum many-body scars. 
\end{abstract}
\maketitle
\section{Introduction}
\label{sec:Intro}

Systems that violate the eigenstate thermalization hypothesis (ETH) have been intensely studied in recent years ~\cite{Moudgalya_Review,Moudgalya_HSF,Serbyn_QMBS_Review,Papic_QMBS_Review,Chandran_QMBS_Review}. ETH is a set of constraints on the eigenstates of a quantum time-evolution operator such that  for (almost) any initial state, observables evolve to a steady state at late times and exhibit only those correlations dictated by thermal equilibrium at the initial energy density. Such systems in effect ``act as their own heat bath''. Generic isolated many-body systems are believed to satisfy ETH, giving an intuitive justification for applying various standard techniques of equilibrium statistical mechanics to time evolution. ETH can be violated by the presence of an extensive set of conservation laws, as occurs in integrable models~\cite{Deutsch_ETH_Review}. Similar phenomena are believed to emerge in many-body-localized systems as a consequence of strong disorder~\cite{Nandkishore_MBL_ETH}. In both cases, ETH is expected to be strongly violated, meaning that at most a set of measure zero of the eigenstates of such a quantum model obeys ETH.


A distinct symmetry-related mechanism for breaking ETH is Hilbert-space fragmentation (HSF)~\cite{Sala_HSF,Rakovszky_SLIOM,Moudgalya_HSF,Moudgalya_Numerics,Khudorozhkov_RecentHSF}. HSF occurs when the dimension of the commutant algebra of a Hamiltonian grows exponentially with system size, ``fragmenting'' the Hilbert space into exponentially many dynamically disconnected subspaces. The commutant algebra is defined for a {family} of Hamiltonians of the form $H =  \sum_j J_j h_j$ as the set of all operators that commute with {\em each} of the local few-body operators $h_j$, and hence with $H$ for any choice of the $J_j$.  Strong HSF occurs when the size of each subspace is a vanishing fraction of the total size, leading to a strong violation of ETH. Weak HSF (and consequently a weak ETH violation) occurs when almost all eigenstates belong to a single subspace, i.e.\ the collective measure of all other subspaces tends to zero. 


In this paper, we introduce an SU($M$)-symmetric bipartite spin model which exhibits HSF even in the presence of random and long-range couplings.  Half of the $2N$ spins are in the fundamental ${\mathbf{M}}$ representation of SU($M$), and half in the conjugate $\overline{\mathbf{M}}$. The Hamiltonian 
consists of a sum over projectors onto the spin-singlet state for each ${\mathbf{M}},\overline{\mathbf{M}}$ pair. 
This Hamiltonian generalizes the nearest-neighbor spin chains of \cite{Batchelor_BQisTL,Aufgebauer_2010}, the most famous example being the ``biquadratic" spin-1 chain. 

We show by explicit construction that our Hamiltonian has a nullspace whose dimension grows exponentially with increasing system size. States in this nullspace are frustration-free, i.e.\ are annihilated by each of the projectors. When all the coefficients of the projectors are positive (the ``ferromagnetic'' case), they are ground states. Some of these states are product states, and the remainder can be found by exploiting the fact that the nullspace of the clean model (uniform couplings) is the same. For spin $M\geq3$, we show that the resulting nullspace dimension grows exponentially with increasing system size. We identify precisely the SU($M$) representations in the nullspace and find their dimensions increase only polynomially with $N$. The exponential growth of the nullspace size therefore results in HSF.

This large nullspace arises as a consequence of a non-trivial commutant algebra. This algebra is a subalgebra of the Read-Saleur (RS) algebra~\cite{ReadSaleur}, whose elements commute with all the generators of certain presentations of the Temperley-Lieb algebra, including those used to build the nearest-neighbor analogs of our model~\cite{Moudgalya_HSF}. Remarkably, here we find that some of the HSF and the commutant algebra present in the nearest-neighbor chains survives their generalisation to our model as long as we preserve bipartiteness. 

Our model can be viewed as a bipartite generalization of the Sachdev-Ye (SY) model~\cite{Sachdev_SYmodel,Georges_SY_ResidEntropy}, which has random all-to-all couplings and SU($M$) symmetry. 
The SY model, along with the closely related Sachdev-Ye-Kitaev (SYK) model of quartically coupled Majorana fermions, have received much attention for their links to quantum chaos. Many aspects of their thermalizing dynamics may be accessed using field-theoretic techniques. We also deploy such techniques in the limit of large $M$ and $N$, and find that the resulting equations  are similar to those obtained for the bipartite SYK model~\cite{BipartiteSYK}. They thus indicate similar non-Fermi-liquid behaviour and emergent conformal symmetry at low energies for the appropriate disorder-averaged observables. 
However, our model exhibits the additional feature of HSF not apparent in the disorder-averaged field theory and not believed to be present in the SY model.

We show how to modify our Hamiltonian using a long-range variant of Shiraishi-Mori spectral embedding~\cite{Shiraishi_SMEmbeddingFormalism}. The product states are then spread throughout the spectrum and constitute quantum many-body scars~\cite{Serbyn_QMBS_Review}. The existence of these states results in a weak violation of ETH, and can be arranged so as to lead to periodic revivals, a hallmark of scars in experimental setups. Our model thus provides a rare example of a non-local model that violates ETH via fragmentation or scarring. 






Our paper is organized as follows. In Sec.\  \ref{sec:IntroducingRSSY}, we define our Hamiltonian and demonstrate its key properties, namely the exponential growth of its nullspace with system size and the resulting Hilbert-space fragmentation. In Sec.\ \ref{sec:ExactRSSYresults}, we solve the uniform-coupling version of our model, and use this solution to derive an exact formula for the dimension of the nullspace of the full model. 
We show that the fragmented sectors can be identified solely by the SU($M$) symmetry: states in certain SU($M$) representations are always part of the nullspace, while others never belong to the nullspace.
We find the commutant algebra in Sec.\ \ref{sec:ReadSaleur}, and show how it is a remnant of the Read-Saleur algebra.
In Sec.~\ref{sec:FieldTheory} we exploit the bipartite Sachdev-Ye structure to discuss the structure of the disorder-averaged correlations of this model at low energies.
We generalize the Shiraishi-Mori embedding formalism to non-local Hamiltonians in Sec. \ref{sec:SMFormalismRSSY}, and show that the resulting scar states violate ETH. We apply this formalism  to yield a non-local model with an exponentially growing number of scar states, leading to weak fragmentation. 
In Sec. \ref{sec:Conclusion}, we provide concluding remarks.

\section{Basic Properties}
\label{sec:IntroducingRSSY}

In this section, we introduce our Hamiltonian $H_M$ and derive some of its key properties. We explain how it can be thought of as a bipartite version of the well-known Sachdev-Ye model, an all-to-all SU($M$)-symmetric model. Despite the resemblance, we show that it has some significantly different properties. Namely, we demonstrate that the dimension of the nullspace of $H_M$ grows exponentially with increasing system size. The states in this nullspace are frustration-free and correspond to ground states for ferromagnetic couplings. A consequence is that our model is Hilbert-space fragmented, with each map between two nullspace eigenstates constituting an element of the commutant algebra. 

\subsection{The Hamiltonian}
\label{subsec:DefRSSY}
We study a spin-$S$ SU($M$)-symmetric bipartite system, with $M=2S+1$. The system is composed of $2N$ spins, with $N$ spins transforming in the fundamental representation $\mathbf{M}$ of SU($M$) and $N$ transforming in the conjugate representation $\overline{\mathbf{M}}$. We will usually 
label the spins of the former by upper indices $i$ and the latter by lower indices $j$ with $1\leq i,j \leq N$. The corresponding states for each spin are labeled respectively as $\ket{a^i}$ and $\ket{a_j}$ with 1$\,\leq a\leq$\,$M$. We work in a basis where the symmetry generators are built from operators acting on a given spin as
\begin{equation}
T^{a,i}_b = \ket{a^i}\bra{b^i},\qquad S^a_{b,j} = -\ket{b_j}\bra{a_j}\ ,
\label{eq:UMgenerators}
\end{equation}
while acting trivially on the others. The operators
\begin{equation}
J^a_b = \sum_{i=1}^N T^{a,i}_b + \sum_{j=1}^N S^a_{b,j}.
\label{eq:globalSUM}
\end{equation}
generate a $U(M)$ algebra, but the $U(1)$ generator $\sum_{a} J^a_a$ vanishes on any state in our Hilbert space. Thus effectively the symmetry is SU($M$).

We study the bipartite model with all two-spin interactions between each pair ($i$,$j$) with $i$ in the $\mathbf{M}$ half and $j$  in the $\MM$ half. The spins in each half do not interact amongst themselves. The SU($M$)-invariant two-spin interaction can be written in terms of the (unnormalized) projector  onto a singlet, which is given in this basis by
\begin{align}
P^{i}_{j} &= -\sum_{a,b=1}^M T^{a,i}_b S^b_{a,j} = \sum_{a,b=1}^M \ket{a^ia_j}\bra{b^ib_j}\ .
\label{Pdef}
\end{align}
The eigenstate of $P^{i}_{j}$ with non-vanishing eigenvalue $M$ is
\begin{equation}
\ket{\psi^{0,i}_{j}} = \sum_{a=1}^M \ket{a^ia_j}\ .
\label{eq:singletstate}
\end{equation}
The most general such Hermitian Hamiltonian is comprised of a sum over singlet projectors for each pair with arbitrary real coefficients  $r^i_j$:
\begin{equation}
    H_M = \sum_{i,j=1}^{N} r^i_{j} P^{i}_{j}\ .
\label{eq:defRSSY}
\end{equation}

For $M$\,=\,2, the projector is the usual spin-$\tfrac12$ Heisenberg interaction.
When $M$\,=\,3, a familiar way to rewrite the Hamiltonian is in terms of the ``biquadratic'' spin-1 interaction. Namely,
we let $\boldsymbol{S}^{i} = ((S^x)^i, (S^y)^i, (S^z)^i)$ denote the usual spin-1 SU(2) generators acting on spin $i$, and likewise for $\boldsymbol{S}_{j}$ acting on spin $j$. The projector then can be unitarily transformed into
\begin{equation}
\begin{aligned}    
U_3 P_i^jU_3^{\dagger} &= (\boldsymbol{S}^{i} \cdot \boldsymbol{S}_{j})^{2} - \mathbb{1},\\
\hbox{where }\ U_3&=\prod_{j=1}^N \exp(i\pi S^z_j)\exp(i\pi S^x_j)\ .
\end{aligned}
\end{equation}
As has long been known \cite{Batchelor_BQisTL}, the symmetry of the biquadratic spin chain is enhanced to SU(3) with generators given by \eqref{eq:globalSUM}. 

The nearest-neighbor uniform chain analog of $H_M$ is found by setting  $r_i^j$\,=\,0 unless $j=i,i+1$, yielding a chain with $2N$ sites and nearest-neighbor interactions:
\begin{equation}
    H_{\rm nn}= \sum_{i=1}^{N} \big(r_{i}^i P^{i}_{i}+r^i_{i+1} P^{i}_{i+1}\big)\ .
\label{Hnn}
\end{equation}
This Hamiltonian 
has long been studied (see e.g.\ \cite{Batchelor_BQisTL,Aufgebauer_2010}), because its generators obey the Temperley-Lieb algebra. With uniform couplings, the model is integrable, and many properties can be computed utilising the XXZ chain whose generators satisfy the same algebra.

Remarkably, $H_{\rm nn}$ has a large commutant algebra whose generators were derived by Read and Saleur (RS)~\cite{ReadSaleur}. Each element of  this algebra commutes with each projector from \eqref{Pdef} {\em individually}, and so commutes with $H_{\rm nn}$ for any couplings. For $M$\,=\,2, where $H_{\rm nn}$ is the Heisenberg chain, the commutant algebra is simply SU(2). However, for $M$\,$\geq$\,3, the dimension of the RS algebra grows exponentially with system size, leading to exponentially large degeneracies in the spectrum. A heuristic way of understanding why is to note that the spectrum in essence follows from properties of the Temperley-Lieb algebra, while the dimensions of the Hilbert spaces of the XXZ chain and $H_{\rm nn}$ grow as $2^{2N}$ and $M^{2N}$ respectively. Thus for $M$\,$\geq$\,3, it is natural to expect large degeneracies in the spectrum of the latter. We discuss the ensuing RS algebra in more detail in Sec.~\ref{sec:ReadSaleur}.

\subsection{Hilbert-space fragmentation}
\label{subsec:HSFinRSSY}

One of central results of our paper is that $H_M$ with completely arbitrary couplings $r_i^j$ has for $M\ge 3$ has a nullspace whose dimension grows exponentially with $N$, and as a consequence the model exhibits Hilbert-space fragmentation. We show these facts here by a rather elementary analysis. We derive an exact formula for the dimension of this nullspace in Sec.~\ref{sec:ExactRSSYresults}, and show how the SU($M$) symmetry allows us to precisely characterize the corresponding states. In Sec.~\ref{sec:ReadSaleur}, we connect this result to the survival of a non-trivial subalgebra of the RS algebra.

Since $H_M$ from Eq.~\eqref{eq:defRSSY} is a sum of projectors, its nullspace $\mathcal{N}$ for arbitrary couplings consists of those states annihilated by each projector $P^{i}_j$ individually. Such states are said to be frustration-free. 
An exponentially large number of states in $\mathcal{N}$ are product states, and so can be found easily. We write a product state $\ket{\psi_{\mathcal{N}}}$ as
\begin{equation}
\begin{aligned}
    \ket{\psi_{\mathcal{N}}}= 
    \otimes\ket{t^1 t^2 \ldots t^N s_1 s_2 \ldots s_N},
\end{aligned}
\end{equation}
where each $t^i,s_j\in\{1,\ldots,M\}$ denotes the state of the spin $i,j$ in the ${\mathbf{M}}$ and $\overline{\mathbf{M}}$ irreps respectively.
As apparent from \eqref{Pdef}, if we choose $t^i$ and $s_j$ such that
\begin{equation}
    t^i\leq z<s_j,\, \, \, \text{for any}\, z\in\{ 1,\ldots, M-1\}
\label{tsconstraint}
\end{equation}
then $\ket{\psi_{\mathcal{N}}}$ will be annihilated by each projector in Eq.~\eqref{eq:defRSSY}. 
The total number of possible values for all the levels $t^i$ and $s_j$ such that the above constraint is satisfied is $(z(M-z))^N$ for each value of $z$. 


Obviously, for $M\geq 3$, the number of product states in the nullspace obeying \eqref{tsconstraint} grows exponentially with $N$.  
Since all these product states belong to non-trivial SU($M$) multiplets, the number of them provides a lower bound to the dimension $D_{\mathcal N}$ of the nullspace:
\begin{align}
D_{\mathcal N} \ge \sum_{z=1}^{M-1} z^N(M-z)^N\ 
\end{align}
For $N$ large, this bound is sharply peaked around $z\approx M/2$, so we expect the exponential growth to be 
\begin{equation}
    D_\mathcal{N} \propto \begin{cases} \big(\tfrac{M}2\big)^{2N} & M\ \text{even}, \\ \big(\tfrac{M^2-1}4\big)^N  , & M\ \text {odd}.\end{cases}
    \label{DNpower}
\end{equation}
For ferromagnetic couplings ($r^i_j>0$ for all $i$ and $j$), each state in the nullspace  
is a ground state, yielding an exponentially growing ground-state degeneracy.

For any couplings, the large nullspace results in Hilbert-space fragmentation. 
In its analysis we follow the approach of ~\cite{Moudgalya_HSF,Moudgalya_Standard,Moudgalya_QMBS}. The bond algebra $\mathcal{A}$ of $H_M$ is generated by the projectors $P^{i}_j$ and the identity operator,
with elements given by linear combinations of arbitrary products of generators. 
The commutant algebra $\mathcal{C}$ is the algebra of operators $\widehat{O}$ 
that commute with each generator of $\mathcal{A}$ individually:
\begin{equation}
    \mathcal{C}=\left\{\widehat{O}:\left[P^{i}_j, \widehat{O}\right]=0,\, \text{for} \, 1\leq i,j\leq N \right\}.
\end{equation}
Under $\mathcal{C}$, the Hilbert space splits into dynamically disconnected Krylov subspaces. The total number $K$ of Krylov subspaces is bounded  by the dimension of $\mathcal{C}$~\cite{Moudgalya_HSF} as
\begin{equation}
    \sqrt{\dim (\mathcal{C})} \leq K \leq \dim (\mathcal{C}).
\end{equation}

Hilbert-space fragmentation occurs in local models when $\log(\dim(\mathcal{C}))$, and hence $\log(K)$, grows as a volume law with system size. The fragmentation is non-trivial when $\mathcal{A}$ is non-abelian, as in our case. (If $\mathcal{A}$ were abelian, the model would be trivially solvable.) As $H_M$ is effectively zero-dimensional, we define fragmentation to occur when $\log(\dim(\mathcal{C}))$ grows linearly with $N$. We also require that this exponential growth does not arise solely from the degeneracies arising from the SU($M$) symmetry, i.e.\ the number of SU($M$) multiplets in the nullspace must grow exponentially.

Since the states in the nullspace $\mathcal{N}$ are annihilated by each $P^{i}_j$ individually, maps between them that leave all other states invariant are elements of $\mathcal{C}$. Indeed, for any two states $\ket{\psi_A},\ket{\psi_B} \in \mathcal{N}$, we have that
\begin{equation}
    \Big[P^{i}_j,\,\ket{\psi_A}\bra{\psi_B}\Big] = 0,\, \, \, \text{for}\, 1\leq i,j\leq N.
\end{equation}
Hence, $\dim(\mathcal{C})\geq(\dim(\mathcal{N}))^2$, and so the size of the commutant algebra grows exponentially with $N$. This leads to Hilbert-space fragmentation, with each state in $\mathcal{N}$ constituting its own one-dimensional Krylov subspace.

In section \ref{sec:HSF2} we derive an exact formula for the dimension of the nullspace, and find precisely which SU($M$) representations are present in it. This analysis allows us to show that the degeneracies are exponentially larger than those arising solely from the SU($M$) symmetry. Thus our model exhibits Hilbert-space fragmentation.

\subsection{Comparing $H_M$ to the Sachdev-Ye Model}
\label{subsec:RSSYvsSY}

As an SU($M$)-symmetric spin model with disordered long-range interactions, $H_M$ is reminiscent of the well-known Sachdev-Ye (SY) model~\cite{Sachdev_SYmodel,Georges_SY_ResidEntropy}. The SY model is a disordered all-to-all variation on the SU($M$)-symmetric Heisenberg model~\cite{Sachdev_SYmodel}. Its Hamiltonian is given by
\begin{equation}
    H_{\rm SY}= \frac{1}{\sqrt{MN}}\sum_{i < j }^{N} \sum_{a,b=1}^M g_{ij} \mathcal{S}^a_{b,i}  \mathcal{S}^b_{a,j},
\label{eq:SYdef}
\end{equation}
where $\mathcal{S}^a_{b,i}$ are generators of SU($M$) acting non-trivially at site $i$, and the $g_{ij}$ are independent and identically distributed Gaussian random variables with vanishing means and variances that do not scale with $N$ or $M$. 

The crucial distinction between the SY model and ours is that all sites in the former transform in the same SU($M$) irreducible representation. When that representation is the fundamental $\mathbf{M}$, each operator $\mathcal{S}^a_{b,i}  \mathcal{S}^b_{a,j}$ is equal (up to a constant shift) to the permutation operator $\mathbb{P}_{ij}$ swapping the spins on sites $i$ and $j$.

As a consequence,  $H_{\rm SY}$ and $H_M$ have rather different physical properties. The permutation operators $\mathbb{P}^{ij}$ in SY split the $M^2$ states into $M(M+1)/2$ and $M(M-1)/2$-dimensional subspaces.
The singlet-projector operators $P^{i}_j$ comprising $H_M$ have rank 1,  annihilating all but one of the basis states in the $M^2$-dimensional Hilbert space formed by two spins. It is therefore not surprising that the $H_M$ has a non-trivial nullspace, unlike SY. We would also expect the ranks of the projectors $P^{i}_j$ to influence the rest of the spectrum, with  $H_M$ having a higher density of states close to energy zero than this SY model. We have checked this behavior numerically and indeed confirmed it at small $N$.

The SY model is therefore expected to obey ETH, as tested in the fermionic analog, the Sachdev-Ye-Kitaev (SYK) model \cite{Sonner_ETHSYK1,Haque_ETHSYK2}. A reason is that when all sites are in the fundamental representation, the SY model has the same bond algebra as the SU($M$) Heisenberg model. Indeed any $\mathbb{P}_{ij}$ is the product of nearest-neighbour permutation operators, e.g.\ for $j>i$, $\mathbb{P}_{ij}=\mathbb{P}_{i,i+1}\mathbb{P}_{i+1,i+2}\cdots\mathbb{P}_{j-1,j}$~\cite{Sagan_symmetric}. 
The same bond algebra implies the same commutant algebra as well, and including random all-to-all couplings will not increase the symmetry. The SU($M$) Heisenberg model for $M>$\,2 is not integrable and has no known symmetry algebra beyond SU($M$). It therefore obeys ETH, and SY model thus should as well.

\section{Exact results}
\label{sec:ExactRSSYresults}

We here obtain the nullspace $\mathcal{N}$ of $H_M$. As $\mathcal{N}$ is defined as the set of states annihilated by all the projectors, it is independent of the couplings. We thus can find it  by solving the ``clean'' case, where all couplings are the same. We derive an exact formula for its exponentially large dimension, and show precisely which SU($M$) representations appear in $\mathcal{N}$. 

\subsection{The energies of the clean model}
\label{subsec:DiagonalisingRSSY}

The clean version of $H_M$ is given by setting all couplings $r^i_j$ in Eq.~\eqref{eq:defRSSY} to one:
\begin{equation}
    H_{\text{clean}} = \sum_{i,j=1}^{N} P^{i}_{j} = -\sum_{i,j=1}^{N}\sum_{a,b=1}^M T^{a,i}_b S^b_{a,j}\ .
\end{equation}
This Hamiltonian is invariant under permutations of the $N$ spins in the $\mathbf{M}$ representation amongst themselves, and likewise for those in the $\MM$. 
This symmetry enhancement allows the spectrum of $H_{\rm clean}$ to be computed from a purely group-theoretical analysis. 

The method for solving $H_{\rm clean}$ is to write it as a sum of Casimir invariants. Any representation of a Lie algebra possesses a quadratic Casimir operator that commutes with all the generators of the algebra in that representation. The full system we are studying is comprised of $N$ spins in the $\mathbf{M}$ representation of SU($M$) and $N$ in the $\MM$, i.e.\ the representation is $\mathbf{M}^{\otimes N} \otimes \MM^{\otimes N}$ with generators given in (\ref{eq:globalSUM},\ref{eq:UMgenerators}). The corresponding Casimir operator is 
\begin{align} 
C_{\mathbf{M}^{\otimes N}\otimes\MM^{\otimes N}} = \sum_{a,b=1}^M J^a_b J^b_{a}.\end{align}
We also need the SU($M$) Casimir operators for just the $\mathbf{M}$ spins and for just the $\MM$ spins, which are
\begin{equation}
\begin{aligned} 
C^{}_{\mathbf{M}^{\otimes N}} &= \sum_{a,b=1}^M \mathcal{T}^{a}_b \mathcal{T}^{b}_{a} = -\frac{N^2}{M}+  \sum_{i=1}^N\sum_{a,b=1}^M {T}^{a,i}_b T^{b,i}_{a}\,,\\
 C_{\MM^{\otimes N}} &=\sum_{a,b=1}^M \mathcal{S}^{a}_{b} \mathcal{S}^b_{a,j}= -\frac{N^2}{M}+ \sum_{a,b=1}^M \sum_{j=1}^N {S}^{a}_{b,j} {S}^b_{a,j}\,,
\end{aligned}
\end{equation}
where the SU($M$) generators acting on the $\mathbf{M}^{\otimes N}$ and $\MM^{\otimes N}$ subspaces are respectively
\begin{align*}
\mathcal{T}^{a}_b = \sum_{i=1}^N \Big(T^{a,i}_b - \tfrac1{M} \delta_b^a\Big),\quad
\mathcal{S}^{a}_b = \sum_{j=1}^N \Big(S^{a}_{b,j} + \tfrac1{M} \delta_b^a\Big),
\end{align*} 
giving traceless operators as needed. 

The clean Hamiltonian therefore can be written in terms of these three Casimir operators as
\begin{equation}
    H_{\text{clean}} =  \tfrac{N^2}{M}+\tfrac12 \big(C^{}_{\mathbf{M}^{\otimes N}} + C_{\MM^{\otimes N}} - C_{\mathbf{M}^{\otimes N}\otimes \MM^{\otimes N}}\big).
\label{Hclean}
\end{equation}
Other models consisting of a sum of SU$(M)$ Casimirs have been analyzed \cite{Beverland_CasHamil1,Perlin_CasHamil2,Jakab_CasHamil3,Jakab_Bipartite_SU3_Hamil}, but all spins in these models are in the $\mathbf{M}$ representation.

The key property of the quadratic Casimir operator is that for an irreducible representation it is proportional to the identity. Irreducible representations of SU($M$) are characterized by an $M$-dimensional vector  ${\rho}=[\rho_1,\rho_2,\ldots,\rho_M]$ comprised of integers satisfying
\begin{equation}
\rho_1\geq\rho_2\geq\ldots\geq\rho_M\geq0.
\label{repcond}
\end{equation} 
Each such vector  can be written as a partition $P_{S,k}$ of $S=\sum_a \rho_a$ into $k$ positive integers, where $k\le M$. For any partition with $k<M$, we set $\lambda_b$\,=\,0 for $b>k$. 
The vector corresponding to the fundamental $\mathbf{M}$ representation is [1,0,0,$\dots$0], while for the conjugate $\MM$ it is [1,1,$\dots$,1,0]. SU($M$) 
representations related by shifting all $\rho_a\to \rho_a-1$ (until $\rho_M=0$) are equivalent.

The quadratic Casimir operator obeys 
\begin{align}
C_\rho = c_\rho \mathbb{1}\quad \hbox{ for }\rho\hbox{ irreducible.}
\label{Cc}
\end{align}
For any representation of a semisimple Lie algebra, the number $c_\rho$ can be determined by standard techniques. For SU($M$), it is \cite{Perelomov_UMCasimir}
\begin{align}
c_\rho = \sum_a\Big( \rho_a^2 \ + \ (M+1-2a) \rho_a\Big) - \frac{1}{M}\Big(\sum_a\rho_a\Big)^2 .
\label{cdef}
\end{align}
where here and everywhere unlabeled sums over $a$ run from 1 to $M$. The Casimir ${c}_\rho$ is indeed invariant under the shift $\rho_a\to \rho_a-1$.

Group theory alone yields all the eigenvalues of $H_{\rm clean}$.
Finding them requires decomposing the Hilbert space $\mathbf{M}^{\otimes N}\otimes\MM^{\otimes N}$ into irreducible representations of SU$(M)$. Since \eqref{Hclean} involves also the Casimir operators for the two types of spins alone, we must also keep track of the representations of SU($M$) in the corresponding subspaces. 
The simplest way to understand tensor products of SU($M$) representations is via Young diagrams. The Young diagram for an irreducible representation $\rho$ is a collection of boxes glued together such that the $a$th row has $\rho_a$ boxes in it.
To take the tensor product of two representations, one glues the boxes from the corresponding Young diagrams together subject to certain rules (see e.g.\  \cite{Hamermesh_SYTisPermutationIrrep} or any text on Lie algebras).


We first need to decompose the tensor products $\mathbf{M}^{\otimes N}$ and $\MM^{\otimes N}$ into a direct sum over irreducible representations $\lambda$ and $\overline{\gamma}$ respectively, and then decompose the tensor product of each $\lambda$ and $\overline{\gamma}$ pair. 
The irreducible representations $\lambda$ in the decomposition of $\mathbf{M}^{\otimes N}$ obey $\sum_a\lambda_a = N$. They therefore can be written as integer partitions of $N$ into at most $M$ parts: 
\begin{align}
\mathbf{M}^{\otimes N} =\oplus m(\lambda)\,\lambda,\qquad \lambda\in P_{N,k}\hbox{ for } k \le M.
\label{decomp1}
\end{align}
where $m(\lambda)$ is the ``multiplicity'', the number of times $\lambda$ appears in this direct sum. We give its value below.
The decomposition of $\MM^{\otimes N}$ can be characterized similarly by taking advantage of conjugate representations. 
The conjugate  $\overline{\gamma}$ is defined via $\overline{\gamma}_a = {\gamma_1}-\gamma_{M+1-a}$. The conjugate of a tensor product is the tensor product of the conjugates. We then utilize \eqref{decomp1} to get
\begin{align} 
\MM^{\otimes N} 
=\oplus m(\gamma)\,\overline{\gamma},\quad \gamma\in P_{N,l}\hbox{ for } l \le M.
\label{decomp2}
\end{align}
Here $\sum_a\overline{\gamma}_a=M\gamma_1-N$.

To decompose the full Hilbert space into irreducible representations, we then take the tensor product of each pair to give $\lambda\otimes\overline{\gamma} = \oplus \rho$.
The Hilbert space thus can be decomposed into sectors labeled by $(\lambda,\overline{\gamma},\rho)$. The clean Hamiltonian is diagonal in such a basis, with any state $\ket{\psi_{\lambda,\overline{\gamma},\rho}}$ in a given sector obeying
\begin{align}
H_{\rm clean} \ket{\psi_{\lambda,\overline{\gamma},\rho}} = \Big(\tfrac{N^2}{M}+\tfrac12\big(c_\lambda+c_{\overline{\gamma}}- c_\rho\big)\Big) 
\ket{\psi_{\lambda,\overline{\gamma},\rho}}\ .
\label{Hcleanpsi}
\end{align}
This spectrum of the clean model is highly degenerate, and below 
we compute its ground-state degeneracy.

\subsection{The ground states of the clean model}

To find the dimension of the nullspace of $H_{M}$, we first construct the states annihilated by $H_{\rm clean}$.  Since eigenvalues of $H_{\rm clean}$ and Casimirs of SU($M$) are always non-negative, from \eqref{Hcleanpsi} we have
\begin{align}
c_\rho \le 2\tfrac{N^2}{M} + c_\lambda+c_{\overline{\gamma}},
\label{cinequality}
\end{align}
with ground states given by those representations $\rho \in \lambda\otimes\overline{\gamma}$ satisfying the equality. 

Thus for a given pair $\lambda,\overline{\gamma}$, candidates for ground states must maximize $c_\rho$.  To find this maximum, we exploit a useful inequality for SU($M$) representations: for any $\rho\in\lambda\otimes \overline{\gamma}$, then \cite{Schlosser_sun_formula}
\begin{equation}
     \sum_{a=1}^b\big(\lambda_a + \overline{\gamma}_a\big) \geq \sum_{a=1}^b \rho_a 
\label{sumident}    
\end{equation}
for any $b$. It follows from the explicit expression \eqref{cdef} that this $c_\rho$ is maximized when $\rho$ obeys the equality in \eqref{sumident}, i.e.\ ${\rho}={\lambda}+{\overline{\gamma}}$. Here the Young diagram for $\rho$ is given by gluing those for $\lambda$ and $\overline{\gamma}$ together horizontally.

The zero-energy ground states of $H_{\rm clean}$ are therefore those $\bl$ and $\bog$ for which
\begin{align}
c_{\bl+\bog}=2\tfrac{N^2}{M} + c_\bl+c_{\,\bog}.
\label{czero}
\end{align}
By construction $\sum_a\bl_a$\,=\,$N$ and $\sum_a\bog_a$\,=\,$M\bg_1- N$, so that \eqref{cdef} yields
\begin{equation}
\begin{aligned}
c_{\bl +\bog}-c_\bl-c_{\,\bog}&=2\sum_a\bl_a\bog_a-2N\bg_1 + 2\tfrac{N^2}{M}\\
&= -2\sum_a\bl_a\bg^{}_{M+1-a} + 2\tfrac{N^2}{M}\ .
\end{aligned}
\end{equation}
The condition \eqref{czero} then reduces to
\begin{align}
\sum_a\bl_a\bg^{}_{M+1-a} = 0\ .
\label{zerocond}
\end{align}
In (\ref{decomp1},\ref{decomp2}) we defined $\lambda$ and $\gamma$ as partitions of $N$. Since both $\bl_a$ and $\bg_{M+1-a}$ are non-negative, one of these two must vanish for all $a$ for \eqref{zerocond} to be satisfied. 

The condition \eqref{zerocond} therefore is equivalent to the remarkably simple constraint $k+l \le M$. We thus find that the ground states of $H_{\rm clean}$ and the null space of $H_M$ are all those $\ket{\psi_{\bl,\bog,\bl+\bog}}$ { with } 
\begin{align}
\boxed{\ \bl\in P_{N,k},\quad \bg\in P_{N,l},\quad  k+l \le  M. \ }
\label{gs}
\end{align}
This constraint is both necessary and sufficient for obtaining a zero-energy ground state of $H_{\rm clean}$.

We thus have proved that the Hilbert space of $H_M$ is fragmented: all states satisfying \eqref{gs} are part of the null space. Since there are many such SU($M$) representations, this degeneracy goes well beyond the consequences of this global symmetry. In section \ref{sec:HSF2}, we prove an even stronger statement: any state in a representation $\rho$ that obeys $\rho=\bl+\bog$ is in the nullspace, as long as $\bl$ and $\bg$ satisfy the conditions of \eqref{gs}. 

\subsection{Exact dimension of the nullspace}
\label{subsec:ExactRSSYDegeneracies}

We here present an exact formula for the dimension of the nullspace $\mathcal{N}$ of $H_M$ as a function of $M$ and $N$. As discussed in section \ref{subsec:HSFinRSSY}, all states in the nullspace are annihilated by each projector $P^{i}_j$. Because $H_{\rm clean}$ is a sum over these projection operators with positive coefficients, its eigenvalues are non-negative, and all eigenstates with zero eigenvalue are annihilated by all $P^i_j$.  Hence the null states for any $r^i_j$ are given by the zero-energy ground states of the clean model. We therefore may make use of the solution of the latter to derive the nullspace dimension $D_{\mathcal{N}}$. 
It therefore follows from purely group-theoretical calculations.

The nullspace is spanned by all the states in \eqref{gs}. The multiplicities $m(\bl)$ and $m(\bg)$ in (\ref{decomp1},\ref{decomp2}) give rise to one kind of degeneracy: any of the copies of $\bl$ and $\bog$ can be used to make $\rho=\bl+\bog$. The second kind of degeneracy comes from the fact that any of the states in the $\rho$ representation give rise to a ground state. The number of these is $d(\bl+\bog)$, the dimension of this representation. The resulting number of ground states, and hence the dimension of the null space $\mathcal{N}$, is then
\begin{equation}
    D_{\mathcal{N}} = \sum_{\substack{k,\ell=1 \\ k+\ell\leq M}}^{M-1}\  \sum_{\bl\in P_{N,k}}\ \sum_{\bg\in P_{N,\ell}} m(\bl)m(\bg)d(\bl+\bog),
    \label{eq:gsdegenN}
\end{equation}
which indicates that this degeneracy is significantly greater than can be explained via any obvious symmetry of the model.

The multiplicities and dimensions of these representations follow from standard group-theory methods. The $N$ spins in the $\mathbf{M}$ representation are exchanged amongst each other by elements of the permutation group $S_N$, which in the clean model is a symmetry. The multiplicities $m(\bl)$ in \eqref{decomp1} are given by the representation of the corresponding irreducible representation of this $S_N$. Both this representation and the SU$(M)$ representation $\bl$ are labelled by the same Young diagram, a fact known as Schur-Weyl duality. The dimension of an $S_N$ representation is given by 
\begin{equation}
    m(\bl)  = \frac{N!}{h(\bl)},\qquad h(\bl)\equiv\prod_{a,b} h_{a, b}
    \label{eq:dS}
\end{equation}
where $h_{a,b}$ denotes the ``hook length'' of the box at row $a$, column $b$ of the Young diagram for $\bl$, and the product in $h(\bl)$ is over all boxes. A hook comprises the box $(a,b)$, all the boxes to the right of it in the same row $a$, and all the boxes below it in the same column $b$. The hook length $h_{a,b}$ is the number of such boxes.
The dimension $d(\rho)$ of a SU($M$) representation also
involves the product of hook lengths. It is 
\begin{equation}
\begin{aligned}
    d(\rho) = \frac{1}{h(\rho)} \prod_{a=1}^{M}\frac{(\rho_a+M-a)!}
    {(M-a)!}
    \label{eq:dR}
\end{aligned}
\end{equation}


For $M$\,=\,2, the formula \eqref{eq:gsdegenN} exhibits no great surprises. Here both $k$ and $l$ must be 1, and the only partition $P_{N,1}$ is $[N,0]$, which has $m([N,0])$\,=\,1. The corresponding SU(2) representation has dimension $d([2N,0])=2N$\,+\,1, so we find $D_{\mathcal{N}}=2N$\,+\,1, the dimension of the usual ferromagnetic multiplet. 

As we noted in \eqref{DNpower}, for $M\ge 3$ the nullspace dimension grows exponentially with $N$. For $M$\,=\,3 the precise asymptotic is
\begin{equation}
    D_{\mathcal{N}}\Big|_{M=3} \approx \frac{9}{4}\,2^{N}N^2\ \hbox{ for large }N.
\label{D3}
\end{equation}
To prove \eqref{D3}, we first note that using the explicit expressions (\ref{eq:dS},\ref{eq:dR}) for the multiplicities and the dimension in the exact formula \eqref{eq:gsdegenN} for $M$\,=\,3 yields \begin{equation}\begin{aligned}
\label{D3sum}
&D_{\mathcal N}\Big|_{M=3}
= (N+1)^3 + \sum_{n=1}^{\lfloor N/2 \rfloor}\mathcal{P}_n\,\binom{N}{n},\\[0.1cm]
&\mathcal{P}_n=(1+N-2n)^2(2+2N-n)\frac{1+N+n}{1+N-n}\ .
\end{aligned}
\end{equation}
We define $\alpha$ via $n=N/2 - \alpha N$, so that the binomial is sharply peaked at $\alpha$ small. We take $N$ large and $\alpha$ small, and use Stirling's formula to approximate it as 
\[
\binom{N}{\tfrac{N}{2}-\alpha N} =
2^N \sqrt{\frac{2}{N\pi}} e^{-2\alpha^2 N}\Big(1+\mathcal{O}\big(\alpha^2,\alpha^4 N,N^{-1}\big)\Big).\]
This expression is exponentially small unless $\alpha$ is order $1/\sqrt{N}$ or smaller, justifying neglecting the other terms. We then can approximate 
\begin{align} \mathcal{P}_n=18\,\alpha^2 N^3 +\mathcal{O}\big(\alpha N^2,\alpha^3N^3\big)\label{Pnapprox}\end{align}
and replace the sum over $n$ in \eqref{D3sum} with an integral over $Nd\alpha$, giving
\[D_{\mathcal N}\big|_{M=3} \approx 2^N \cdot 
18 N^3 \sqrt{\frac{2N}{\pi}} \int_0^{\frac12}
d\alpha\, \alpha^2\, e^{-2N\alpha^2}
\ .\]
Since the integrand is sharply peaked around $\alpha=0$, we can extend the upper limit of integration to $\infty$ and then do the Gaussian integral, yielding \eqref{D3}. The corrections are suppressed by $1/N$: the terms in \eqref{Pnapprox} suppressed by only $1/\sqrt{N}$ cancel after doing the integral.


Another limit of Eq.~\eqref{eq:gsdegenN} worth considering is the large $M$ limit for fixed $N$. Since the rank of each projector $P^{i}_j$ remains fixed at 1, and the total number of projectors is fixed at $N^2$, we expect the nullspace to dominate in the large $M$ limit. This is indeed the case: we show in App.~\ref{app:LimLargeS} that the limiting form is the dimension of the total Hilbert space:
\begin{equation}
    \lim_{M\to\infty} \frac{D_{\mathcal{N}}}{M^{2N}} =1.
\label{Mlarge}    
\end{equation}
We note that this effect does not happen in the SY model with spins in the fundamental representation: the permutation operators $\mathbb{P}^{ij}$ split the Hilbert space of two spins into symmetric and antisymmetric subspaces, which have comparable dimensions in the large $M$ limit.


\subsection{SU($M$) and the nullspace}
\label{sec:HSF2}

We have demonstrated degeneracies in the spectrum of $H_M$ going well beyond SU($M$), as for $M>$\,2 many different representations obey the nullspace condition \eqref{gs}. The number of each such representation is the product of multiplicities $m(\bl)m(\bg)$, and it grows exponentially with $N$. For example, the binomial in the SU(3) formula \eqref{D3sum} arises from the multiplicities, not the degeneracies coming from a particular SU($M$) representation.

However, we see no evidence for degeneracies and fragmentation outside the nullspace. Thus despite its being exponentially large, the nullspace for $H_M$ still comprises a vanishing fraction of the full Hilbert space for $M$ finite. Thus fragmentation occurs only in a set of measure zero of the full Hilbert space, and almost all states obey ETH. 

The SU($M$) representations appearing in the nullspace defined by \eqref{gs} are rather special. The reason is that all the projectors $P^i_j$ must annihilate states in the nullspace, and when forming $\rho\in \lambda\otimes \overline{\gamma}$, no singlets can appear. 

This property leads to a rather interesting feature of $H_M$. We show here that determining whether a given eigenstate is in the nullspace requires only knowing how it transforms under SU($M$). Namely, any eigenstate of $H_M$ in the SU($M$) representation $\br=\bl+\bog$ satisfying the conditions of \eqref{gs} must be in the nullspace. Any eigenstates of $H_M$ in any other representations are not in the null space. 

Precisely, any eigenstate of $H_M$ can be labeled non-uniquely by the triple $\lambda,\gamma,\rho$ where  $\lambda,\gamma\in P_{N,M}$ and $\rho\in \lambda\otimes\overline{\gamma}$, just as shown in \eqref{Hcleanpsi} for $H_{\rm clean}$. In general the eigenvalues  depend on more data than just these representations, but eigenstates in the null space of $H_M$ satisfy the more restrictive properties given in \eqref{gs}. We prove that the only way to have $\rho=\br$ for any such $\br$ is for the corresponding $\lambda=\bl$ and $\gamma=\bg$ as well.

Assume $\bl$ and $\bg$ satisfy Eq. \eqref{gs}, with $\br=\bl+\overline{\bg}$. Say $\br$ also occurs in the decomposition of some other $\lambda\otimes\overline{\gamma}$. In the decomposition, the boxes in the Young diagram for $\lambda$ will be rearranged according to the tensor product rules into rows of length $n_1,\ldots,n_M$, with $\sum_{a=1}^M n_a=N$, as they are attached to the rows in the diagram of $\overline{\gamma}$ to construct $\br$. Row by row in the diagram of $\br$, this yields the set of equations
\begin{equation}
    \bg_1-\bg_{M-a+1}+\bl_a+m = \gamma_1-\gamma_{M-a+1}+n_a,
    \label{eq:EqSet}
\end{equation}
where the integer $m$ accounts for the fact SU($M$) irreps are equivalent up to a constant shift in the lengths of all rows. Summing the left and right hand sides of Eq. \eqref{eq:EqSet} yields $\bg_1+m = \gamma_1$, giving
\begin{equation}
    -\bg_{M-a+1}+\bl_a = -\gamma_{M-a+1}+n_a
    \label{eq:EqSet2}
\end{equation}
for each $a=1\dots M$.
Let $k$ be the number of nonzero elements in $\bl$, and $\ell$ the number of nonzero elements in $\bg$, with $k+\ell\leq M$ by Eq. \eqref{gs}. Summing the first $k$ equations in Eq. \eqref{eq:EqSet2} then yields
\begin{equation}
    N=\sum_{a=1}^k n_a - \sum_{a=1}^k\gamma_{M-a+1}.
\end{equation}
Because the first sum here is at largest $N$ and the second non-negative, we must have $\sum_{a=1}^k n_a=N$ and so $n_a=0$ for $a\geq k+1$. Using this result in Eq. \eqref{eq:EqSet2} yields $m=0$, $\gamma=\bg$, and $n_a=\bl_a$ for all values of $a$.

The final step comes from noting that the tensor product rules do not allow boxes in a Young diagram to move upward from their original row. Hence the Young diagram of $\lambda$ must have $k$ rows or fewer, and the boxes in these rows can only be moved downward in the construction of the Young diagram of $\br$ to obtain $n_a=\bl_a$ for $a\leq k$. However, since $\overline{\gamma}_a=\overline{\bg}_a=\overline{\bg}_1$ for $a\leq k$, any such downward movement would not be consistent with the tensor product rules: boxes originally in the same row cannot end up in the same column after taking the tensor product. (The reason is that tensor indices are symmetrized in the rows of Young diagram, antisymmetrized in the columns.) Thus the only consistent possibility for $\rho=\bl+\bog\in\lambda\otimes\overline{\gamma}$ satisfying \eqref{gs} is to have both $\gamma$\,=\,$\bg$ and $\lambda$\,=\,$\bl$, as claimed.

\section{The Read-Saleur symmetry algebra and its remnant}
\label{sec:ReadSaleur}


A remarkable feature of the nearest-neighbor chain \eqref{Hnn} is the presence of a large commutant algebra discovered by Read and Saleur (RS) \cite{ReadSaleur}. The resulting degeneracies occur throughout almost all the spectrum. Their number grows exponentially with system size, and so results in Hilbert-space fragmentation \cite{Moudgalya_HSF}. We have shown that the exponentially large dimension of the nullspace and the ensuing fragmentation survive the introduction of the many additional couplings in $H_M$.
As described in section \ref{subsec:HSFinRSSY}, this behavior automatically results in a commutant algebra acting within this nullspace. In this section we demonstrate that this algebra is a subalgebra of the RS algebra.


\subsection{The Read-Saleur algebra}
\label{subsec:ReadSaleurReview}

We here describe the Read-Saleur commutant algebra of the nearest-neighbor chain \eqref{Hnn}. To make the structure more apparent, we renumber the sites as  $p=1\dots 2N$, with odd $p$ labeling spins in the $\mathbf{M}$ representation and even $p$ labeling those in the $\MM$. We also relabel the generators of $H_{\rm nn}$ as
$e_{2i-1}\equiv P^i_i$ and $e_{2i}\equiv P^{i}_{i+1}$ so  
\begin{align}
H_{\rm nn} = \sum_{p=1}^{2N} r_p e_p,\quad e_p = \sum_{a,b=1}^M  \ket{a_p a_{p+1}}\bra{b_pb_{p+1}} \ .
\end{align}
with $r_p$ the non-vanishing random couplings. 
Each projector $e_p$ onto a nearest-neighbour SU($M$) singlet has only one non-vanishing eigenvalue out of the $M^{2}$ possibilities. Since the nearest-neighbor chain has only $2N$ such projectors, one expects large degeneracies for large enough $M$.

Comparison with the XXZ chain suggests that exponentially large degeneracies occur for any $M\ge 3$. The generators $e_p$ satisfy the Temperley-Lieb algebra \cite{Batchelor_BQisTL}:
\begin{align}
\big(e_p\big)^2 = M e_p\ ,\qquad e_p e_{p\pm 1}e_p = e_p
\label{TLdef}
\end{align}
and $e_pe_{p'}=e_{p'}e_p$ with $|p-p'|>1$. Many interesting lattice Hamiltonians are written in terms of the $e_p$; one prominent example is the XXZ chain. In the uniform-coupling case any such chain is integrable, and so the relation between the two spectra can be determined exactly (see \cite{Aufgebauer_2010} and references therein). Since the Hilbert spaces for XXZ and $H_{\rm nn}$ are $2^{2N}$ and $M^{2N}$-dimensional respectively, there must be exponentially large degeneracies in the spectrum of $H_{\rm nn}$ when $M>3$. 

Read and Saleur \cite{ReadSaleur} showed how the degeneracies are independent of the couplings $r_p$ in $H_{\rm nn}$. They did so by constructing the irreducible representations of the Temperley-Lieb algebra in this chain. The resulting degeneracies are completely independent of the couplings, following solely from understanding how the Temperley-Lieb algebra works here. Not surprisingly, the nullspace has the largest degeneracy, but large degeneracies are found for all but a small set of states.



A key tool of Read and Saleur's results is the construction of the commutant algebra $\mathcal{C}^{\text{RS}}$ of the $e_p$. It is generated by $\mathbf{J}^{(s)}$ that satisfy
\begin{align}
e_p\mathbf{J}^{(s)} = \mathbf{J}^{(s)} e_p =0
\label{eJJe}
\end{align}
for all $p$. RS find a linear basis set for the generators in terms of the same building blocks \eqref{eq:UMgenerators} used to construct the SU($M$) generators, namely
\begin{equation}
    J^a_{b,2i-1} = T^{a,i}_{b},\qquad J^a_{b,2j} =  S^a_{b,j}.
\end{equation}
Generators of $\mathcal{C}^{\text{RS}}$ are of the form
\begin{equation}
\begin{aligned}
\mathbf{J}^{(s)} &= \sum_{a_1,b_1,...} \widetilde{g}^{b_1 b_2 \ldots b_s}_{a_1 a_2 \ldots a_s}  \mathbf{J}_{b_{1} b_{2} \ldots b_{s}}^{a_{1} a_{2} \ldots a_{s}},\\ 
\label{eq:RSJdef}
    \mathbf{J}_{b_{1} b_{2} \ldots b_s}^{a_{1} a_{2} \ldots a_s}&\equiv\sum_{1\leq p_1 < p_2 < \ldots < p_s \leq 2N} J_{b_{1},p_1}^{a_{1}} J_{b_{2},p_2}^{a_{2}} \cdots J_{b_{s},p_s}^{a_{s}}
\end{aligned}
\end{equation}
for $0\leq s\leq 2N$. Setting $s$\,=\,1  yields the SU($M$) generators.
For \eqref{eJJe} to be satisfied, the necessary and sufficient conditions on the coefficients are
\begin{equation}
     \sum_{\hat{a}=1}^M {\widetilde{g}}^{\ldots,\hat{a},b_{p+1}\ldots}_{\ldots a_p,\hat{a},\ldots} = \sum_{\hat{a}=1}^M  {\widetilde{g}}^{\ldots,b_{p},\hat{a},\ldots}_{\ldots, \hat{a},a_{p+1},\ldots} = 0 
     \label{eq:gconsRS}
\end{equation}
for all $p$. More explicit expressions for the $\mathbf{J}^{(s)}$ can be found in Ref.~\onlinecite{ReadSaleur}. 

As apparent from \eqref{eq:RSJdef}, the RS algebra includes the universal enveloping algebra $\mathfrak{U}(\hbox{SU}(M))$ (i.e.\ products of the SU($M$) generators in this representation). It is easy to check that the $J^{(s)}$ map between different representations of SU($M$), and so go beyond  $\mathfrak{U}(\hbox{SU}(M)$.  RS derive a variety of interesting properties of $\mathcal{C}^{\rm RS}$, for example showing how gluing together multiple systems allows one to define a tensor-product structure. More generally, they show the algebra is Morita equivalent to $\mathfrak{U}_q(\hbox{sl}(2))$, the commutant algebra of some other Temperley-Lieb systems such as the XXZ chain. However, many interesting questions remain; even the minimal set of generators of $\mathcal{C}^{\rm RS}$ and their commutation relations is not known.


\subsection{The commutant algebra}
\label{subsec:CommutantAlgRSSY}

We have shown explicitly in Sec.~\ref{subsec:ExactRSSYDegeneracies} that degeneracies in the null space survive the inclusion of all $N^2$ couplings in $H_M$, and that one can define elements of the commutant algebra mapping in between them. We show here that these elements form a subalgebra of the RS algebra, in particular of the algebra generated by the $\mathbf{J}^{(2N)}$. 

In the nearest-neighbor case, the $\mathbf{J}^{(2N)}$ act non-trivially on the nullspace of $H_{\rm nn}$ and annihilate all other states. The reason is that the $\mathbf{J}^{(2N)}$ annihilate any state with at least one nearest-neighbour SU($M$) singlet, and the eigenstates of $H_{\rm nn}$ with non-zero eigenvalues are solely composed of such states \cite{ReadSaleur}. Furthermore, for any two states in the nullspace of  $H_{\rm nn}$, there is an operator amongst the $\mathbf{J}^{(2N)}$ that maps between them. 

We generalize this result here. Since all nullstates of $H_M$ are also nullstates of $H_{\rm nn}$, maps between the former are given by operators $\mathbf{J}^{(2N)}$ that satisfy certain additional constraints.  For notational ease, we relabel the indices as
\begin{equation}
\mathbf{J}^{(2N)} = \sum {g^{}}^{b^1 \cdots b^{N}d_1 \cdots d_{N}}_{a^1 \cdots a^{N} c_1 \cdots c_{N}}\,  T^{a_1,1}_{b_1} \cdots T^{a^{N}\!,N}_{b^{N}} S^{c_1}_{d_1,1} \cdots S^{c_{N}}_{d_{N}\!,N}.
\label{eq:RSSYJdef}
\end{equation}
where this sum and all those in this section are over the appropriate $a^i,b^i,c_j,d_j$ from 1 to $M$. To be in the commutant algebra of $H_M$, these operators obey
\begin{equation}
    \mathbf{J}^{(2N)} P^{i}_{j} = P^{i}_{j}\, \mathbf{J}^{(2N)} = 0,\quad \hbox{for }\,  1\leq i,j\leq N.
    \label{eq:RSSYgcons}
\end{equation}
These constraints require that the coefficients satisfy
\begin{align}
     \sum_{\{b\},\{c\}} \delta^{c_i}_{b^j}\,{g}^{\,\cdots b^{j}\cdots\cdots\cdots}_{\,\cdots \cdots \cdots c_{i}\cdots} =  \sum_{\{a\},\{d\}} \delta^{a^i}_{d_j}\,{g}^{\,\cdots \cdots\cdots d_{j} \cdots}_{\,\cdots  a^{i}\cdots\cdots\cdots} = 0
\label{eq:RSSYgconstraints}
\end{align}
for all $1\leq i,j\leq N$.

A general null state of $H_M$ is given by
\begin{equation}
    \ket{\psi}=\sum_{\{a\},\{d\}} {\psi}_{a^1  \cdots a^{N}}^{d_1 \cdots d_{N}} \ket{a^1\cdots a^{N}d_1 \cdots d_{N}},
    \label{eq:GeneralState}
\end{equation}
where the $a^i$ and $d_j$  transform in the $\mathbf{M}$ and $\overline{\mathbf{M}}$ representations respectively, and the coefficients satisfy
\begin{align}
    \sum_{\{a\},\{d\}}   \delta^{a^i}_{d_j} {\psi}_{a^1\cdots a^{N}}^{d_1\cdots d_{N}} =  0,\, \hbox{ for } \, 1\leq i,j\leq N.
    \label{eq:RSSYgs}
\end{align}
The resemblance to the constraints on $\mathbf{J}$ is not coincidental. We can find a solution to \eqref{eq:RSSYgconstraints} by taking any two solutions $\psi,\chi$ of \eqref{eq:RSSYgs} and defining 
\begin{equation}
    {g^{}}^{b^1 \cdots b^{N}d_1 \cdots d_{N}}_{a^1 \cdots a^{N} c_1 \cdots c_{N}} = \psi_{a^1 a^2\cdots a^{N}}^{d_1 d_2 \cdots d_{N}}\Big(\chi_{b^1 \cdots b^{N}}^{c_1 \cdots c_{N}}\Big)^*.
\end{equation}
We thus have proved that $\ketbra{\psi}{\chi}$ can be written as an operator $\mathbf{J}^{(2N)}$ from Eq.\eqref{eq:RSJdef}.


Regarding  the other energy levels of $H_M$, our numerics found no sign of degeneracies other than those arising from the global SU($M$) symmetry.  This suggests that the only operators $\mathbf{J}^{(s)}$ which act nontrivially on eigenstates outside the nullspace are elements of $\mathfrak{U}$(SU($M$)). This suggestion is in harmony with Read and Saleur's derivation of the symmetry generators for the nearest-neighbor chain \cite{ReadSaleur}. Their construction utilizes a ``dimer" basis for the Hilbert space, where each state is written as a product of spin-singlet (dimer) states.  Operators in the Temperley-Lieb algebra act non-trivially on the dimers, whereas those in the RS algebra leave the dimers untouched. The non-local projectors $P^i_j$ in $H_M$, however, mix the dimer states in a much more complicated way, and so there is no reason to expect degeneracies beyond SU($M$). Indeed, from this point of view it is rather surprising that even the non-trivial nullspace survives.

\section{Low-energy field-theory description}
\label{sec:FieldTheory}

In Sec.~\ref{subsec:RSSYvsSY} we discussed how $H_M$ resembles a bipartite version of the SY model but with Hilbert-space fragmentation. Here we sharpen the comparison by analyzing a field-theory limit using manipulations familiar from studies of the SY/SYK models~\cite{Sachdev_SYmodel, ChowdhuryReview}.  We will take the limit of a large number of spins and of large $M$, and subsequently the limit of frequencies small compared to microscopic scales when we explicitly discuss constraints on conformal \textit{ansatze} for the low-energy correlations.
As we explain next, we also need to generalize the spins to larger representations of SU($M$).

\subsection{Spins in antisymmetric representations}

A natural generalization of $H_M$ is to take spins in representations other than the fundamental ones. We utilize the antisymmetric representation $\mathbf{A}_m$ of SU($M$), which has a Young diagram with one column and $m$ rows. We consider $N$ spins in the $\mathbf{A}_m$ representation and $\widetilde{N}$ in the $\overline{\mathbf{A}}_m$\,=\,$\mathbf{A}_{M-m}$ representation. 
The Hamiltonian is the obvious generalization of \eqref{eq:defRSSY}, namely
\begin{align}
H_{M,m} = \frac{1}{\sqrt{MN}}\sum_{i=1}^N\sum_{j=1}^{\tilde{N}} g^i_j P^i_j
\label{HMm}
\end{align}
where the projectors are as in \eqref{Pdef} with the sum running instead to $d_m=\binom{M}{m}$, the number of states per site. We have included random couplings $g^i_j$ and allow for $N\ne\tilde{N}$. 

The generalized Hamiltonian $H_{M,m}$ remains a sum of positive semidefinite operators, so the nullspace still consists of those states annihilated by each two-site term individually. The calculation of the nullspace dimension thus proceeds as in Sec.~\ref{sec:ExactRSSYresults}. The clean model Hamiltonian is written in terms of Casimirs as
\[
    H_{M,m}^{\rm clean} = \tfrac{ m^2 \big(N^2+\tilde{N}^2\big)}{2M}+\tfrac12 \Big(C^{}_{\mathbf{A}_m^{\otimes N}} + C_{\overline{\mathbf{A}}_{m}^{\otimes \tilde{N}}} - C_{\mathbf{A}_m^{\otimes N}\otimes \overline{\mathbf{A}}_{m}^{\otimes \tilde{N}}}\Big)
\]
just like \eqref{Hcleanpsi}. The tensor products can be decomposed into irreducible representations $\lambda$\,$\in$\,$\mathbf{A}_m^{\otimes N}$, $\overline{\gamma}$\,$\in$\,${\overline{\mathbf{A}}}_{m}^{\otimes \tilde{N}}$ and $\rho\in\gamma\otimes\lambda$. One then finds as in \eqref{gs} that the ground states are states where $\rho=\lambda+\overline{\gamma}$ and $k+l\le M$, with Young diagrams for $\lambda$ and $\gamma$ having $k$ and $l$ rows respectively.

For $m\ne M/2$, similar arguments to those for $m$\,=\,1 show that the dimension of the nullspace grows exponentially with $N$. 
For $m=M/2$, the model is equivalent to the SY model with spins in the representation $A_{M/2}$. We thus expect no Hilbert-space fragmentation. Indeed, all but one of the possible $\lambda$ have more than $M/2$ rows,  and likewise for $\gamma$. Hence there is only one SU($M$) representation in the nullspace, that with a rectangular Young diagram with $m=M/2$ rows and $N+\tilde{N}$ columns. It has multiplicity one in the tensor products, so fragmentation does not occur.

\subsection{Field-theory description at large $M,m$}

Here we utilize field-theory techniques \cite{Sachdev_SYmodel, ChowdhuryReview} to show how some of the physics closely resembles that of the SY and SYK models. We study the Hamiltonian $H_{M,m}$ from \eqref{HMm} with spins in antisymmetric representations. Here we set $\tilde{N} /N = b$, and $m/M = c$ with $b$ and $c$ order 1. 

We represent the spins in the $A_m$ and $\overline{A}_m$ representations using fermionic operators $f^i, \tilde{f}_j$:
\begin{equation}
\begin{aligned}\label{eq:Sf}
    \mathcal{S}^{a,i}_{b} &= f^{a,i\dagger} f^i_{b}; \quad \sum_{a= 1}^M f^{a,i\dagger} f^i_{a} = m,\\
    \tilde{\mathcal{S}}^a_{b,j} &=  \tilde{f}^{a\dagger}_{j} \tilde{f}_{b,j}; \quad \sum_{a= 1}^M \tilde{f}^{a\dagger}_{j} \tilde{f}_{a,j} = M-m,
\end{aligned}
\end{equation}
where the fermions obey standard anticommutation relations: $\{f^{a,i}, f^{a,i\dagger}\} = \{\tilde f_{b,j}, \tilde f^\dagger_{b,j}\} = 1$, and all other anticommutators vanish. The Hamiltonian \eqref{HMm} written in terms of spins contains interactions of the SY form between the two groups of spins
\begin{equation}\label{eq:RSSYGen_m}
    H = \frac{1}{\sqrt{MN}}\sum_{i = 1}^{N} \sum_{j = 1}^{\tilde{N}} \sum_{a,b=1}^M g^i_j \mathcal{S}^{a,i}_{b} \tilde{\mathcal{S}}^{b}_{a,j}.
\end{equation}
The coefficients $g^i_j$ are independent and identically distributed Gaussian random variables with $\overline{g^i_{j}} = 0$ and $ \overline{(g^i_{j})^2} = g^2$. 

We introduce the on-site (and diagonal in SU($M$) index) Green's functions at $T = 0$:
\begin{equation}
    G(\tau) = - T_\tau \langle f^{a,i}(\tau) f^{a,i\dagger}(0) \rangle, \  \tilde G(\tau) = - T_\tau \langle \tilde f_{a,j}(\tau) \tilde f_{a,j}^\dagger(0) \rangle.
\end{equation}
That we calculate disorder-averaged quantities is implicit throughout this section; whether the quantities of interest are well-represented by their disorder averages is a separate question, and one which we shall not confront. One can calculate these Green's functions by writing the Hamiltonian of Eq.~\eqref{eq:RSSYGen_m} in terms of fermions using Eq.~\eqref{eq:Sf}, with chemical potentials $\mu, \tilde{\mu}$ to fix the respective fillings. We then observe that the diagrams which do not vanish in the large $N, M$ limit are of the melon type, just like in the SYK model. The Schwinger-Dyson equations are
\begin{equation}\label{eq:Gdefinition}
    G(i \omega) = \frac{1}{i\omega +\mu- \Sigma(i\omega)},  \quad \tilde G(i\omega) = \frac{1}{i\omega +\tilde{\mu}- \tilde\Sigma(i\omega)},
\end{equation}
with
\begin{equation}\label{eq:Sigmamelon}
    \Sigma(\tau) = \feynmandiagram[inline=(a.base), horizontal=a to b]{
a--[charged scalar]b,
b--[charged scalar, half left]a,
a--[fermion, half left]b
};
=  - U^2 G(\tau)\tilde G(\tau) \tilde G(-\tau),  
\end{equation}
where $U^2 = bg^2/2$ and
\begin{equation}\label{eq:Sigmatildemelon}
    \quad \tilde \Sigma(\tau) = 
    \feynmandiagram[inline=(a.base),horizontal=a to b]{
a--[fermion]b,
b--[fermion, half left]a,
a--[charged scalar, half left]b
};   
=  -\tilde{U}^2 \tilde G(\tau) G(\tau) G(-\tau),
\end{equation}
where $\tilde U ^2 = g^2/2$. In the above diagrams, solid and dashed lines denote the propagators $G$ and $\tilde G$. The disorder average which couples the two vertices is implicit. Finally, the filling constraints of Eqs.~\eqref{eq:Sf} translate to
\begin{equation}\label{eq:Gfilling}
    G(\tau = 0^-) = c ,\quad  \tilde G(\tau = 0^-) =  1-c.
\end{equation} 
All other propagators are smaller by factors of $N, M$ relative to these, and so will be ignored. 
Equations with a similar structure have appeared in the solution of a bipartite Majorana SYK model in Ref.~\onlinecite{BipartiteSYK} and an all-to-all $t$-$J$ model in Ref.~\onlinecite{ChristosPNAS}.

Eqs.~\eqref{eq:Gdefinition} through~\eqref{eq:Gfilling} in general need to be solved numerically. However, in looking for solutions in the $\omega \to 0$ (consequently $|\tau|\to \infty$) limit some further progress can be made. One starts by guessing SYK-like Green's functions in the upper half frequency plane ($\text{Im } z > 0$):
\begin{equation}\label{eq:Gconformalomega}
    G(z) = C \frac{e^{-i (\pi \Delta + \theta)}}{z^{1-2\Delta}}, \quad  \tilde G(z) = \tilde C \frac{e^{-i (\pi \tilde\Delta + \tilde\theta)}}{z^{1-2\tilde\Delta}}.
\end{equation}
These allow us to find the corresponding spectral functions, which yield, via Laplace transforms,  $G(\tau)$ and $\tilde G(\tau)$:
\begin{equation}\label{eq:Gconformaltau}
    G(\tau) = 
    \begin{cases}
        -\frac{C \Gamma(2\Delta)}{\pi} \frac{\sin(\pi \Delta + \theta)}{\tau^{2 \Delta}} & \tau > 0\\
        \frac{C \Gamma(2\Delta)}{\pi} \frac{\sin(\pi \Delta - \theta)}{|\tau|^{2 \Delta}} & \tau < 0,
    \end{cases}
\end{equation}
and an identical expression with tildes for $\tilde G(\tau)$. Defining
\begin{equation}
    \alpha_{\pm} = \frac{C\Gamma(2\Delta)}{\pi} \sin(\pi \Delta \pm \theta)
\end{equation}
(and similar for $\tilde\alpha_\pm$) Eqs.~\eqref{eq:Sigmamelon} and~\eqref{eq:Sigmatildemelon} become, respectively,
\begin{gather}
\Sigma(\tau) = \begin{cases}
    -U^2  \alpha_+ \tilde\alpha_+ \tilde\alpha_-  {|\tau|^{-2\Delta - 4 \tilde \Delta}}    & \tau > 0 \\
    U^2 \alpha_- \tilde\alpha_+ \tilde\alpha_-  {|\tau|^{-2\Delta - 4 \tilde \Delta}}     \tau < 0
   \end{cases}
\end{gather}
and its tilded counterpart.

Again using the inverse Laplace transforms as an intermediate object, we can find $\Sigma(i\omega)$ and $\tilde\Sigma(i\omega)$. Then, assuming that the small-frequency behaviour of the denominators in Eq.~\eqref{eq:Gdefinition} is dominated by the self energies, equating the powers of $\omega$ in the same equations gives $1-2\Delta = 2\Delta + 4 \tilde \Delta - 1$, i.e.
\begin{equation}
    \Delta +\tilde\Delta = \frac{1}{2},
\end{equation}
and equating the coefficients then gives 
\begin{gather}\label{eq:1C}
   - \frac{1}{C} = \frac{\Gamma(2 \Delta) \Gamma^2(2 \tilde\Delta)}{2\pi^3\Gamma(2\Delta +4\tilde\Delta)} C \tilde{C}^2 U^2 (\cos 2\pi \tilde \Delta - \cos 2\tilde \theta)
\end{gather}
and the tilded version. Dividing these two equations, one arrives at
\begin{equation}\label{eq:ratiodelta}
    \frac{\tilde U^2}{U^2} = \frac{1}{b} = \frac{2 \Delta}{1-2\Delta}
    \frac{\cos2\tilde\theta + \cos 2\pi\Delta}{\cos2\theta - \cos 2\pi\Delta}.
\end{equation}
Setting $\theta = \tilde\theta = 0$, which enforces particle-hole symmetry, one recovers the equation for $\Delta$ in the bipartite SYK (i.e., quartic-in-Majoranas) model studied in Ref.~\onlinecite{BipartiteSYK}. Finally, we have equations that fix the fillings in terms of the parameters that enter the Green's functions ~\cite{NotesComplexSYK, ChowdhuryReview}:
\begin{gather}
   \label{eq:luttinger1}  c = \frac{1}{2} - \frac{\theta}{\pi } - \left(\frac{1}{2} - \Delta\right) \frac{\sin 2\theta}{\sin 2\pi\Delta}, 
    \\
   \label{eq:luttinger2} \quad 1-c = \frac{1}{2} - \frac{\tilde\theta}{\pi} - \Delta \frac{\sin 2\tilde \theta}{\sin 2\pi\Delta}.
\end{gather}
Note that these ``Luttinger relations'' cannot be found by substituting the Green's functions of Eq.~\eqref{eq:Gconformaltau} (which are valid only at large $|\tau|$) into Eq.~\eqref{eq:Gfilling} (which concerns the $\tau \to 0$ behaviour of the true Green's functions).
Eqs.~\eqref{eq:ratiodelta},~\eqref{eq:luttinger1},~\eqref{eq:luttinger2} are three equations in the variables $\Delta, \theta, \tilde\theta$, and may be solved numerically. 

We should stress that the conformal \textit{ansatz}, while demonstrably a valid solution to the Schwinger-Dyson equations as $\omega \to 0$, is not guaranteed to be associated with any solution that is valid at higher energies as well. (See, e.g.,  Ref.~\onlinecite{Azeyanagi} for a treatment of this issue in the SYK context.) In this sense the approach described here does not represent a complete solution for the Green's function even in the $N, M \to \infty$ limit. 

We expect that the states in the nullspace, the focus of the rest of this paper, are invisible to the solutions of the Schwinger-Dyson equations discussed here, and would also be invisible to their finite-temperature version  --- even if analytic continuation to real time can be performed. This is because these states are exponentially (in $N$) rare in the spectrum, and so taking the $N\to\infty$ limit (which one has to do for any analytical control) will erase their contribution. While $1/M$ corrections may be possible to treat, and are likely to indicate a spin-glass state \cite{Christos},  the $N\to\infty$ limit must be taken first in such calculations; we are not aware of a way to calculate, say, $1/N$ corrections at large but finite $M$. Furthermore, all the quantities one can calculate in the field-theoretic approach are disorder-averaged. It is not clear \textit{a priori}, even at finite $N$, if disorder-averaged correlation functions should be expected to inherit any non-trivial features from the non-ergodicity of the Hamiltonian.

\section{Scars from the Shiraishi-Mori construction}
\label{sec:SMFormalismRSSY}

We have established that each eigenstate in the nullspace of $H_M$ does not mix with the others under time evolution. As such, it is tempting to label these eigenstates as quantum many-body scars~\cite{Moudgalya_QMBS}. The numerous product eigenstates identified in Sec.~\ref{subsec:HSFinRSSY} have area-law entanglement entropy scaling, a signal of a scar. However, typically a scar is
a rare ETH-violating exception in an otherwise thermal spectrum. Here the situation is rather different, as states in certain SU($M$) representations are in the nullspace, while all others are not. 

We show in this section how to perturb $H_M$ to yield more conventional scars. We utilize the Shiraishi-Mori (SM) embedding formalism \cite{Shiraishi_SMEmbeddingFormalism} to break the SU($M$) symmetry, converting some of the original nullspace eigenstates into quantum many-body scars. To do so, we must address subtleties involved in applying this formalism to non-local models.



A quantum many-body scar for a local Hamiltonian violates ETH. Since global symmetries also constrain time evolution in the obvious fashion, we must split the Hilbert space into sectors $\mathcal{H}_n$, labeled by the corresponding quantum numbers.
Defining $\mathcal{H}_{\text{thermal},n}$ to be largest Krylov subspace of $\mathcal{H}_n$, two conditions are sufficient for an eigenstate $\ket{\psi_{\text{scar}}}$ to be a quantum many-body scar \cite{Moudgalya_QMBS}, namely
\begin{equation}
\begin{aligned}
    \ketbra{\psi_{\text{scar}}}{\psi_{\text{scar}}} \in \mathcal{C}\;,
    \qquad \frac{\dim(\mathcal{H}_{\text{thermal},n})}{\dim( \mathcal{H}_n)}\xrightarrow{N\rightarrow \infty} 1,
    \label{eq:QMBSconds}
\end{aligned}
\end{equation}
where $\mathcal{C}$ is a non-trivial commutant algebra. The first condition ensures that the scar state violates ETH, since it implies that the scar state is an eigenstate of a family of Hamiltonians, whereas ETH-obeying eigenstates generically have a unique parent Hamiltonian ~\cite{Moudgalya_QMBS,Garrison_SingleEstat1,Qi_SingleEstat2}. The second condition ensures that the scar state is a rare exception in a spectrum dominated by thermal eigenstates. We note that $\ket{\psi_{\text{scar}}}$ need not have area-law entanglement entropy~\cite{Langlett_RainbowScars,Moudgalya_QMBS}. Additionally, if the number of scars in a model grows exponentially with $N$, then by Eq.~\eqref{eq:QMBSconds} the dimension of the commutant algebra grows exponentially as well, leading to Hilbert-space fragmentation. The fragmentation in this case is weak, because the second condition of Eq.~\eqref{eq:QMBSconds} says that the spectrum is dominated by one large thermal Krylov subspace.

Following Refs.~\onlinecite{Shiraishi_SMEmbeddingFormalism,Moudgalya_QMBS}, we perturb $H_M$ as
\begin{equation}
    H_{\text{SM}} = \sum_{i,j=1}^{N}  P^{i}_{j} h^{[i]}_{[j]} P^{i}_{j} + H_{\text{pert}},
\label{eq:HAASM}
\end{equation}
where the  $h^{[i]}_{[j]}$ are finite-support operators that are sufficiently general to break the SU($M$) symmetry, and  $H_{\text{pert}}$ is chosen to split the degeneracies between the nullspace eigenstates. Each operator $h^{[i]}_{[j]}$ need not have support on the spins $i$ and $j$, so these indices merely serve as labels. The original Hamiltonian $H_M$ from Eq.~\eqref{eq:defRSSY} corresponds to setting $h^{[i]}_{[j]}=r^i_j \mathbb{1}$ and $H_{\text{pert}}=0$.

Because the SU($M$) symmetry of $H_M$ is broken, the perturbed Hamiltonian $H_{\text{SM}}$ acquires a dominant thermal subspace, and hence satisfies the second condition of Eq.~\eqref{eq:QMBSconds}. Furthermore, when $H_{\text{pert}}$\,=\,0, the nullspaces of $H_{\text{SM}}$ and $H_M$ are the same and so states in it satisfy the first condition of Eq.~\eqref{eq:QMBSconds} as well. However, satisfying these conditions does not automatically guarantee that states in the nullspace  of a non-local model are quantum many-body scars.  This is because  a thermal eigenstate is generally understood to uniquely specify a parent Hamiltonian only for spatially local systems~\cite{Garrison_SingleEstat1,Qi_SingleEstat2}. 
In the absence of locality, we thus cannot conclude that the first condition  is thus is sufficient to imply that an eigenstate violates ETH. 

However, the product eigenstates in the nullspace identified in Sec.~\ref{subsec:HSFinRSSY} have area-law scaling and so violate ETH. Moroever, we can choose $H_{\text{pert}}$ such that the product eigenstates of $H_M$ are split from the remainder of the nullspace. Namely, we perturb by the diagonal on-site U($M$) generators defined in Eq.~\eqref{eq:UMgenerators} via
\begin{equation}
    H_{\text{pert}} = \sum_{\alpha=1}^{M} \left(\sum_{i=1}^{N}c^{\alpha,i} T^{\alpha,i}_\alpha+\sum_{j=1}^{N}d^{\alpha}_j  S^{\alpha}_{\alpha,j}\right),
    \label{eq:Hpert}
\end{equation}
where the $c^{\alpha,i}$ and $d^{\alpha}_j$ are arbitrary distinct real coefficients. The product states from $\mathcal{N}$ remain eigenstates of $H_{\text{SM}}$, and now are situated in quantum-number sectors of $H_{\text{SM}}$ that have a dominant thermal Krylov subspace. They thus satisfy the conditions in Eq.~\eqref{eq:QMBSconds}, and furthermore violate ETH due to their area-law entanglement entropy scaling. The product states hence constitute quantum many-body scars of $H_{\text{SM}}$.
In particular, when the coefficients $c^{\alpha,i}$ and $d^{\alpha}_j$ are rational, any initial state composed of a sum of these product states would exhibit revivals when time evolved with $H_{\text{SM}}$, a hallmark of scars~\cite{Serbyn_QMBS_Review}. Additionally, since the number of such product states grows exponentially with $N$, the perturbed model  $H_{\text{SM}}$ has an exponentially-growing number of quantum many-body scars and is Hilbert space fragmented.

\section{Concluding Remarks}
\label{sec:Conclusion}

The non-local model introduced in this paper possesses a number of remarkable properties. In spite of its all-to-all interactions, the dimension of its nullspace grows exponentially with increasing system size. Furthermore, maps between eigenstates in this nullspace are elements of the commutant algebra. The exponentially growing degeneracy thus leads to Hilbert-space fragmentation in some of the SU($M$) quantum number sectors of the spectrum, violating ETH in a weak sense. Much like the SY model, our model can in appropriate limits be attacked by field-theoretic methods, at least for disorder averaged low-energy observables. However these techniques cannot access the special states in the null space.  The Shiraishi-Mori embedding formalism can be utilised to perturb many of the eigenstates in the nullspace into quantum many-body scars.

Non-local models with Hilbert space fragmentation are rarer in the literature than those with quantum many-body scars. Non-local models have a large bond algebra: for example, the number of generators of the bond algebra of $H_M$ grows quadratically with $N$, as opposed to the linear growth associated with a 1D spin chain. This strongly constrains the dimension of the commutant algebra, making fragmentation less likely.


Previous works have considered how scars and fragmentation can be stable against the addition of longer-range interactions to local Hamiltonians~\cite{Lerose_AlgebraicQMBS,stephen_plaquettesQMBS,Francica_FredkinHSF,
zhang_QMBSinSY,Pakrouski_GroupInvar1,Pakrouski_GroupInvar2}. However, the model studied here is apparently a rare example of one where the non-local terms are the leading contribution to the problem, and cannot be ascribed a notion of spatial distance: it is intrinsically ``zero dimensional''. It may be of interest to explore other problems of that nature which share some features of our model, for instance the model of  Ref.~\onlinecite{Iyoda_WishartSYK} which is also SYK-like and has  extensive degeneracies in its eigenspectrum. We leave this question for the future.


\medskip

\acknowledgments
We thank Sounak Biswas, Maine Christos, and Pablo Sala for helpful discussions. This work was supported by the European Research Council under the European Union Horizon 2020 Research and Innovation Programme via Grant Agreement No. 804213-TMCS (S.A.P.) and by the  Fonds de recherche du Qu\'ebec via Doctoral Level Grant No. 302311 (J.C.-H.). We acknowledge support from EPSRC via grant EP/S020527/1.


\appendix

\section{Limiting form for $M$ large, $N$ fixed}

\label{app:LimLargeS}

In this appendix, we show that for $M$\,$\geq$\,2$N$, the Hilbert space is dominated by the nullspace, as displayed in  \eqref{Mlarge}:
\[
    \lim_{M\to\infty} \frac{D_{\mathcal{N}}}{M^{2N}} = 1 
\]
The proof follows from two facts: 1) for all states in the nullspace as given in \eqref{gs}, the dimension of the corresponding SU($M$) multiplet grows as $M^{2N}$ for fixed $N$, while 2) all other irreducible representations have dimension growing as a lower power of $M$.  

We first note that $k+\ell\leq2N<M$ when $M>2N$. Since the restriction in the sum in \eqref{eq:gsdegenN} is always satisfied, the nullspace dimension simplifies to
\begin{equation}
    D_{\mathcal{N}} = \sum_{k,\ell=1}^{N}\  \sum_{\bl\in P_{N,k}}\ \sum_{\bg\in P_{N,\ell}} m(\bl)m(\bg)d(\bl+\bog).
   \label{eq:dTlargeM}
\end{equation}
The only dependence on $M$ here is through the SU($M$) dimension $d(\br)$. The formula \eqref{eq:dR} for this dimension then can be factored into two pieces by splitting this product into one piece involving the first $k$ rows of the Young diagram for $\br$, and another involving the other rows. The latter part of this diagram can be thought of as a separate Young diagram  with entries $\br^{(M-k)}=[\br_{k+1},\br_{k+2},\ldots, \br_M]=
[\bog_{k+1},\bog_{k+2},\ldots, \bog_M]$. The corresponding contributions of this piece to $d(\br)$ then amount to the dimension $d^{(M-k)}$ of the representation $\br^{(M-k)}$ in the group SU($M-k$), yielding 
\begin{equation}
\begin{aligned}
    d(\br) = d^{(M-k)}\big(\br^{(M-k)}\big) \frac{\prod_{a=1}^{k}(\rho_a+M-a)!}{\prod_{a=1}^k\Big( (M-a)!\prod_{b=1}^{\rho_a} h_{a, b}\Big)}.
    \label{eq:dSUMdecomp}
\end{aligned}
\end{equation}

The resulting large-$M$ behavior is straightforward to determine. We have
\begin{equation}\begin{aligned}
    \prod_{a=1}^{k}\frac{(\br_a+M-a)!}{(M-a)!} =& \prod_{a=1}^{k}\frac{(\bl_a+\bg_1+M-a)!}{(M-a)!} \\&\xrightarrow{M\ \text{large}} M^{N + k\bg_1}\ ,
    \label{eq:Ndep1}
\end{aligned}\end{equation}
where we exploit the facts that $\sum_{a=1}^k\bl_a = N$ and $\bog_a=\bg_1$ for $a\le M-l$.
Because of the latter fact along with $\bg\in P_{N,\ell}$, the first $\gamma_1$ columns in the Young diagram of $\bog$ and hence $\br$ must be at least of length $M-l$. The remaining columns arise only from $\bl$, and so the corresponding hook lengths are independent of $M$, yielding
\begin{align}
\prod_{a=1}^k\prod_{b=1}^{\br_a} h_{a, b} 
\xrightarrow{M\ \text{large}} M^{k\bg_1}h(\bl)
    \label{eq:Ndep2}
\end{align}
Finally, we use the fact the dimensions of an SU($M-k$) representation and its conjugate are the same to derive
\[d^{(M-k)}\big(\br^{(M-k)}\big) = d^{(M-k)}\big(\overline{\br^{(M-k)}}\big)= d^{(M-k)}(\bg)\ .\]
Here the hook lengths are independent of $M$, yielding 
\begin{equation}
\begin{aligned}
   d^{(M-k)}\big(\br^{(M-k)}\big) =& \frac{1}{h(\bg)}
\prod_{a=1}^{\ell}\frac{(\bg_a+(M-k)-a)!}{(M-k-a)!} \\ &\xrightarrow{M\ \text{large}} M^{N}\frac{1}{h(\bg)}\ .
    \label{eq:Ndep3}
\end{aligned}
\end{equation}
Putting the above results together, we see that
\begin{equation}
    d(\bl+\bog) \xrightarrow{M\ \text{large}} \frac{1}{h(\bl)h(\bg)}M^{2N}\ .
\label{dMlarge}
\end{equation}

We proved that states in any SU($M$) representation other than $\br$\,=\,$\bl+\bog$ are not in the nullspace. For $M\ge 2N$, these states must be in representations ${\rho}\in\bl\otimes\bog$ in the tensor product, as there are no restrictions on $\bl\in P_{N,k}$ and $\bg\in P_{N,l}$. Since $\bog_a=\bg_1$ for $a\le M-l$ and $k<M-l$, the inequality \eqref{sumident} means that at least one box in $\bl$ must move downward when forming ${\rho}$.
Moreover, it must move to a row $a$ with $a>M-\ell+1$, as boxes originally in the same row cannot end up in the same column after taking the tensor product. We thus have $\sum_{i=1}^k {\rho}_a < N+k\bg_1$, as opposed to the equality used in the proof of \eqref{eq:Ndep1}. The number of boxes in ${\rho}^{(M-k)}$ must therefore increase, so that   $d^{(M-k)}(\overline{{\rho}^{(M-k)}})$ must decrease relative to \eqref{eq:Ndep3}. Meanwhile, $M$ in \eqref{eq:Ndep2} is unchanged. Thus $d(\rho)$ for $\rho\ne\br$ increases at large $M$ at most with a power $M^{2N-2}$.  

We have shown that the only SU($M$) representations with dimension growing as $M^{2N}$ that appear are those in the nullspace. Since the total dimension of the Hilbert space is $M^{2N}$, the relation
\eqref{Mlarge} must therefore hold.   An amusing by-product of this derivation is the identity
\begin{align}
\sum_{k=0}^N\sum_{\bl\in P_{N,k}} \frac{1}{h(\bl)^2} = \frac{1}{N!}\ .
\label{amusing}
\end{align}
The proof of \eqref{amusing} comes from noting that combining \eqref{dMlarge} and \eqref{Mlarge} yields
\[\sum_{k,l=0}^N\sum_{\bl\in P_{N,k}}\sum_{\bg\in P_{N,l}}  \frac{m(\bl)m(\bg)}{h(\bl)h(\bg)} =1 \]
and then using the first part of  \eqref{eq:dS}.

\bibliography{MyBibliography}

\end{document}